\documentclass[twocolumn,showpacs,preprintnumbers,prd,aps,showkeys]{revtex4}

\usepackage{graphicx, epsfig, epstopdf}
\usepackage{amsmath, amsfonts, amssymb}
\usepackage{bm, array}
\usepackage[usenames]{color}
\usepackage{eurosym}
\usepackage{mdframed}
\usepackage{grffile}
\usepackage{amssymb}
\usepackage{caption}

\def\be{\begin{equation}}
\def\ee{\end{equation}}
\def\beq{\begin{eqnarray}}
\def\eeq{\end{eqnarray}}

\begin{document}

\title{Inflation and Acceleration of the Universe by Nonlinear Magnetic Monopole Fields}
\author{A. \"{O}vg\"{u}n }
\affiliation{Department of Physics, Faculty of Arts and Sciences, Eastern
Mediterranean University, Famagusta, Northern Cyprus}
\email{aovgun@gmail.com}
\date{\today }

\begin{abstract}
Despite impressive phenomenological successes, cosmological models are
incomplete without an understanding of what happened at the big bang
singularity. Maxwell electrodynamics, considered as a source of the
classical Einstein field equations, leads to the singular isotropic
Friedmann solutions. In the context of Friedmann-Robertson-Walker (FRW)
spacetime, we show that singular behavior does not occur for a class of
nonlinear generalizations of the electromagnetic theory for strong fields. A new mathematical model is proposed for
which the analytical nonsingular extension of FRW solutions is obtained by
using the nonlinear magnetic monopole fields.
\end{abstract}

\maketitle

\section{Introduction}

Cosmology has experienced remarkable advances in recent decades as a consequence of tandem observations of type-Ia supernovae and the cosmic microwave background. These observations suggest that cosmological expansion is
accelerating \cite{supernova1}. The last two decades have witnessed enormous
progress in our understanding of the source of this accelerated expansion.
Furthermore, standard cosmology assumes that at the beginning, there must have been an initial singularity -- a breakdown in the geometric structure of space and time -- from which spacetime suddenly started evolving \cite{sing}. The standard cosmological model, with the source of
Maxwell electrodynamics based on the Friedmann-Robertson-Walker (FRW) geometry, leads to a cosmological singularity at a finite time in the past.
In order to solve this puzzle, researchers have proposed many different mechanisms in the literature, such as nonminimal
couplings, a cosmological constant, nonlinear Lagrangians with quadratic terms
in the curvature, scalar inflation fields, modified gravity theories,
quantum gravity effects, and nonlinear electrodynamics without modification of general relativity \cite%
{muk,cos1,cos2,born1,camara,eliz,coupling,novello0,novello1,novello2,vol1,ak3,beck,kruglov1,kruglov2,genci,nonm1,nonm2,horava,ak1,const,novello3,novello4,novello5,ak2}%
.

One possible solution is to explore the evolution while avoiding the cosmic initial
singularity contained in a given nonlinear effect of electromagnetic theory %
\cite{novello0,vol1,kruglov1}. 1934, the nonlinear electrodynamics
Lagrangian known as the Born-Infeld Lagrangian was published by the physicists Max Born and Leopold Infeld \cite{born}. This Lagrangian has the
amusing feature of turning into Maxwell theory for low electromagnetic
fields; moreover, the nonlinear Lagrangian is invariant under the duality
transformation.

To solve the initial singularity problem, the early stages of the universe are assumed to be dominated by the radiation of nonlinear modifications of Maxwell's equations, which include a large amount of electromagnetic and gravitational fields. This is true inasmuch as strong magnetic
fields in the early universe can cause deviations from linear
electrodynamics to nonlinear electrodynamics \cite{born1,camara}. By following recently published procedures \cite{kruglov1,kruglov2}, in this paper the nonlinear magnetic monopole (NMM) fields are
used to show the source of the acceleration of the universe without an initial singularity.

In this paper, we investigate a cosmological model of the universe with NMM
fields coupled to gravity. The structure of the paper is as follows: In
Section II, we briefly introduce NMM fields and consider the universe to be filled by pure nonlinear magnetic fields. In Section III, we
show that the universe accelerates without an initial singularity until it reaches the critical value of the scale factor. In Section IV, we check the classical
stability of the universe under the deceleration phase. In Section V, we report our conclusions.

\section{Nonlinear Magnetic Monopole Fields and a nonsingular FRW Universe}

In cosmology, magnetic fields have become more important since the wealth
of observations of magnetic fields in the universe \cite{w,magneticU1}. Magnetic fields are ubiquitous in understanding the mysteries of the universe.
The action of General Relativity (GR) coupled with NMM fields is given by 
\begin{equation}
S=\int d^{4}x\sqrt{-g}\left[ \frac{M_{Pl}^{2}}{2}R+\alpha \mathcal{L}_{EM}+\mathcal{L}_{NMM}\right],
\label{Ac}
\end{equation}%
where $M_{Pl}$ is the reduced Planck mass, $R$ is the Ricci scalar, and $\alpha$ is the fine-tuning parameter of $\mathcal{L}_{EM}$ Maxwell fields. $\mathcal{L}_{NMM}$ is the Lagrangian of the NMM fields. From a conceptual point of view, this action has the advantage that it does not invoke any unobserved entities such as scalar fields, higher dimensions, or brane worlds. Furthermore, we can ignore the Maxwell fields, ($\alpha=0$), because they are weak compared to the dominant NMM fields in the very early epochs and inflation. However, in the literature there are many proposals of cosmological solutions based on the Maxwell fields plus corrections \cite{novello1,novello2,novello3,novello4,novello5,kruglov1,kruglov2}. Herein, our main aim is to use this method to show that it yields an accelerated expansion phase for the evolution of the universe in the NMM field regime. The new ingredient we add is a modification of the electrodynamics, which has no Maxwell limit.  The
Einstein fields equation and the NMM fields equation are derived from the
action 
\begin{equation}
R_{\mu \nu }-\frac{1}{2}g_{\mu \nu }R=-\kappa ^{2}T_{\mu \nu },  \label{EQ}
\end{equation}

\bigskip where $\kappa ^{-1}=M_{Pl}$, and $\ $%
\begin{equation}
\partial _{\mu }\left( \sqrt{-g}\frac{\partial \mathcal{L}_{NMM}}{\partial 
\mathcal{F}}F^{\mu \nu }\right) =0.
\end{equation}
Note that Maxwell invariant is $\mathcal{F}=F_{\mu \nu }F^{\mu \nu
}=(B^{2}-E^{2})/2>0$, and $F_{\mu \nu }$ is the field strength tensor. 
The magnetic field two-form is $F=Psin(\theta )^{2}d\theta \wedge
d\phi $ or $F_{\theta \phi }=Psin(\theta )^{2}$ where $P
$ is the magnetic monopole charge. Furthermore, it is noted that
in the weak field limit the NMM Lagrangian does not yield the linear Maxwell
Lagrangian \cite{NNM}.
In this work, following a standard procedure, we consider the pure magnetic
field under the following NMM fields Lagrangian
suggested in Ref. \cite{NNM}

\begin{equation}
\mathcal{L}_{NMM}=-\frac{6}{l^{2}\left( 1+\left( {\frac{\beta }{\mathcal{F}}}%
\right) ^{3/4}\right) ^{2}}\,
\end{equation}%
where the $\beta $ and $l$ are the positive
constants. The constant parameter $\beta 
$ will be fixed according to other parameters. The NMM fields Lagrangian is  folded into the homogeneous and isotropic FRW spacetime 
\begin{equation}
ds^{2}=-dt^{2}+a(t)^{2}(dx^{2}+dy^{2}+dz^{2})  \label{frwmetric}
\end{equation}%
or it can be written as follows,%
\begin{equation}
ds^{2}=-dt^{2}+a(t)^{2}\left[ dr^{2}+r^{2}\left( d\theta ^{2}+sin(\theta
)^{2}d\phi ^{2}\right) \right] 
\end{equation}%
where $a$ is a scale factor, to investigate the effects on the acceleration of
the universe. 

The energy momentum tensor%
\begin{equation}
T^{\mu \nu }=K^{\mu \lambda }F_{\lambda }^{\nu }-g^{\mu \nu }\mathcal{L}_{NMM}
\end{equation}%
with%
\begin{equation}
K^{\mu \lambda }=\frac{\partial \mathcal{L}_{NMM}}{\partial \mathcal{F}}F^{\mu
\lambda }
\end{equation}%
can be used to obtain the general form of the energy density $\rho $ and the
pressure $\ p$ by varying the action as following 
\begin{equation}
\rho =-\mathcal{L}_{NMM}+E^{2}\frac{\partial \mathcal{L}_{NMM}}{\partial \mathcal{F}}
\label{rho}
\end{equation}%
and 
\begin{equation}
p=\mathcal{L}_{NMM}-\frac{\left( 2B^{2}-E^{2}\right) }{3}\frac{\partial \mathcal{L}_{NMM}%
}{\partial \mathcal{F}}.  \label{p}
\end{equation}%
Here, it is assumed that the curvature is much larger than the
wavelength of the electromagnetic waves, because the electromagnetic fields
are the stochastic background. The average of the EM fields that are sources
in GR have been used to obtain the isotropic FRW spacetime \cite{tolman}.
For this reason, one uses the average values of the EM fields as follows 
\begin{equation}
\langle \mathbf{E}\rangle =\langle \mathbf{B}\rangle =0,\text{ }\langle
E_{i}B_{j}\rangle =0,
\end{equation}%
\begin{equation*}
\langle E_{i}E_{j}\rangle =\frac{1}{3}E^{2}g_{ij},\text{ }\langle
B_{i}B_{j}\rangle =\frac{1}{3}B^{2}g_{ij}.
\end{equation*}%
Note that later we omit the averaging brackets $\langle $ $\rangle $
for simplicity. The most interesting case of this method occurs only when the average of the magnetic field is not zero. \cite{tolman}. The universe has a magnetic property that the magnetic field is frozen in the
cosmology where the charged primordial plasma screens the electric field. It
is noted that for the pure nonlinear magnetic monopole case, it is clear
that $E^{2}=0.$ Then Eqs. (\ref{rho})  and (\ref{p}) reduce to the simple following form:
\begin{equation}
\rho =-\mathcal{L}_{NMM}  \label{rhoo}
\end{equation}

\bigskip and%
\begin{equation}
p=\mathcal{L}_{NMM}-\frac{2B^{2}}{3}\frac{\partial \mathcal{L}_{NMM}}{\partial \mathcal{F%
}}.  \label{p0}
\end{equation}%
Then the\ FRW metric given \ in Eq.(\ref{frwmetric}) is used to obtain
Friedmann's equation as follows:
\begin{equation}
3\frac{\ddot{a}}{a}=-\frac{\kappa ^{2}}{2}\left( \rho +3p\right),
\label{Feqn}
\end{equation}%
where "." over the $a$ denotes the derivatives with respect to the cosmic
time. The most important condition for the accelerated universe is $\rho
+3p<0$. Here, the NMM field is used as the main source of gravity. Using Eqs. (\ref{rho}) and (\ref{p}), it is found that 
\begin{equation}
\rho +3p=2\mathcal{L}_{NMM}-2B^{2}\frac{\partial \mathcal{L}_{NMM}}{\partial \mathcal{F}}%
.  \label{100}
\end{equation}%
\begin{equation}
=\frac{12\,\left( 2^{7/4}\left( {\frac{\beta }{{B}^{2}}}\right)
^{3/4}-1\right) }{{l}^{2}\left( 1+{2}^{3/4}\left( {\frac{\beta }{{B}^{2}}}%
\right) ^{3/4}\right) ^{3}}.
\end{equation}%
Thus, the requirement $\rho +3p<0$ for the accelerating universe is satisfied
at ( $\left( {\frac{\beta }{{B}^{2}}}\right) ^{3/4}$ $<\frac{1}{2^{7/4}}$), where there is a strong magnetic monopole field in the early stages of the universe to force it to accelerate. By using the conservation of the
energy-momentum tensor, 
\begin{equation}
\nabla ^{\mu }T_{\mu \nu }=0,
\end{equation}%
for the FRW metric given in Eq.(\ref{frwmetric}), it is found that 
\begin{equation}
\dot{\rho}+3\frac{\dot{a}}{a}\left( \rho +p\right) =0.  \label{11}
\end{equation}%
Replacing $\rho $ and $p$ from Eqs. (\ref{rhoo}) and (\ref{p0}), and
integrating, the evolution of the magnetic field under the change of the scale
factor is obtained as follows: 
\begin{equation}
B(t)=\frac{B_{0}}{a(t)^{2}}.  \label{evv}
\end{equation}%
Then, by using Eqs. (\ref{rhoo}) and (\ref{p0}), the energy density $\rho 
$ and the pressure $\ p$ can be written in the form of 
\begin{equation}
\rho =\frac{6}{\,{l}^{2}\left( \Phi \right) ^{2}},~~  \label{rr}
\end{equation}%
\begin{equation}
~~p=-\frac{6}{\,{l}^{2}\left( \Phi \right) ^{2}}+\frac{12\,{a}^{4}{2}%
^{3/4}\beta }{{l}^{2}{\mathit{B}}^{2}\left( \Phi \right) ^{3}}{\frac{1}{\sqrt%
[4]{{\frac{\beta \,{a}^{4}}{{\mathit{B}}^{2}}}}}},  \label{pp}
\end{equation}%
where 
\begin{equation}
\Phi =1+{2}^{3/4}\left( {\frac{\beta \,{a}^{4}}{{\mathit{B}}^{2}}}\right)
^{3/4}.
\end{equation}%
Note that from Eqs. (\ref{rr}) and (\ref{pp}), we obtain the energy density $\rho $ and the pressure $p$, but there is no singularity point
at $a(t)\rightarrow 0$ and $a(t)\rightarrow \infty $. Hence, one finds that, as shown in Fig. (1), 
\begin{equation}
\lim_{a(t)\rightarrow 0}\rho (t)=\frac{6}{l^{2}},~~\lim_{a(t)\rightarrow
0}p(t)=-\frac{6}{l^{2}},  \label{15}
\end{equation}%
\begin{equation}
~~\lim_{a(t)\rightarrow \infty }\rho (t)=\lim_{a(t)\rightarrow \infty
}p(t)=0.  \label{166}
\end{equation}%
From Eqs. (\ref{15}) and (\ref{166}), it is concluded that the energy
density $\rho $ is equal to the negative of the pressure\ $\ p$ ($\rho =-p)$
at the beginning of the universe ($a=0$), similarly to a model of the $%
\Lambda $CDM. The absence of singularities is also shown in the literature (\cite%
{kruglov1,kruglov2}) by using a different model of nonlinear
electrodynamics.

\begin{figure}[tph]
\centering
\includegraphics[width=8cm]{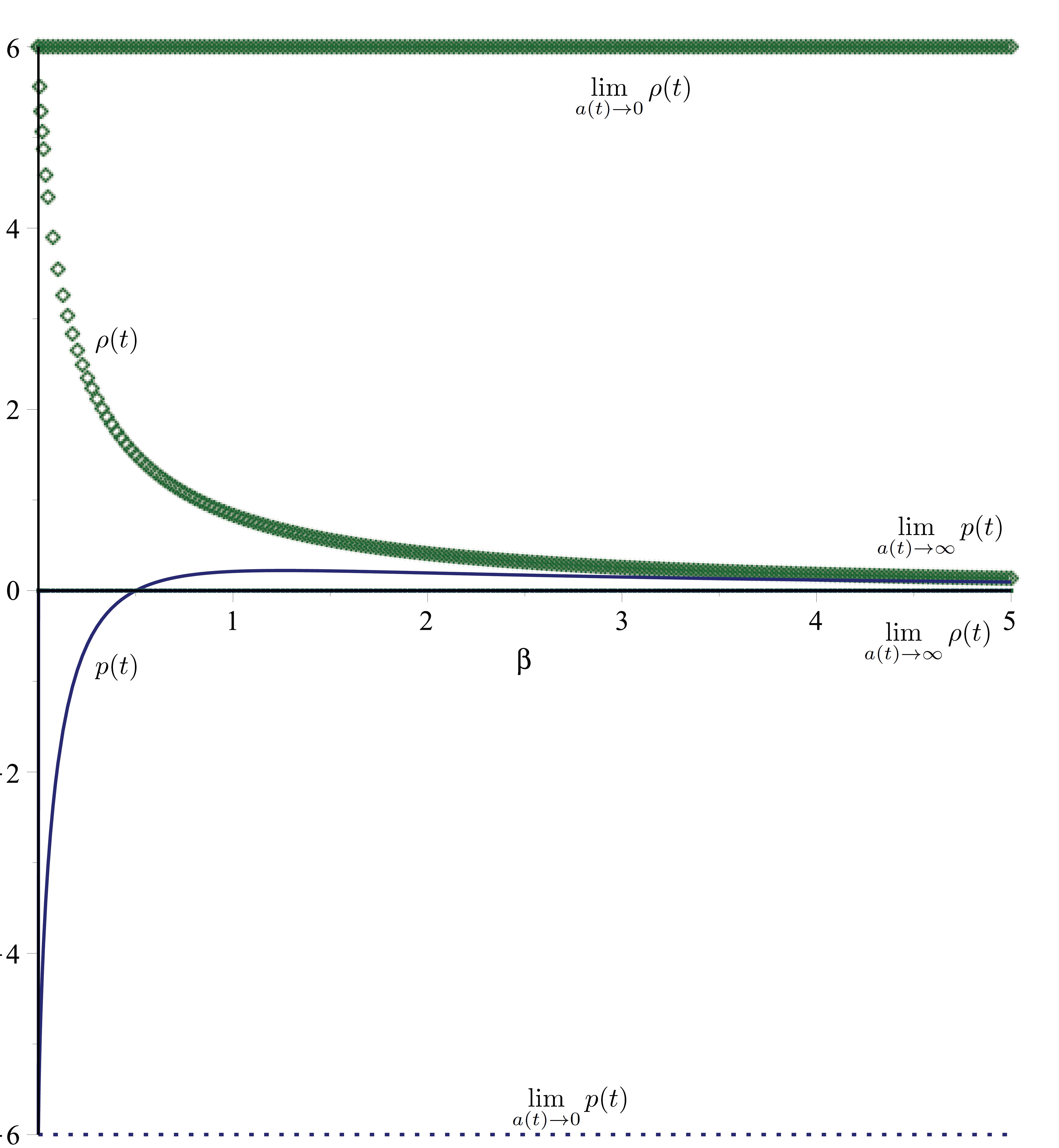}
\caption{plot of the energy density $\protect\rho $ and the pressure $p$
versus $\protect\beta $ (for the cases of $a=1,0,\infty $)}
\end{figure}
\bigskip\ The Ricci scalar, which represents the curvature of spacetime, is calculated by using Einstein's field equation (\ref{EQ}) and the
energy-momentum tensor 
\begin{equation}
R=\kappa ^{2}(\rho -3p).  \label{16r}
\end{equation}%
The Ricci tensor squared $R_{\mu \nu }R^{\mu \nu }$and the Kretschmann
scalar $R_{\mu \nu \alpha \beta }R^{\mu \nu \alpha \beta }$ are also
obtained as 
\begin{equation}
R_{\mu \nu }R^{\mu \nu }=\kappa ^{4}\left( \rho ^{2}+3p^{2}\right) ,
\label{rrrr}
\end{equation}%
\begin{equation}
R_{\mu \nu \alpha \beta }R^{\mu \nu \alpha \beta }=\kappa ^{4}\left( \frac{5%
}{3}\rho ^{2}+2\rho p+3p^{2}\right) .
\end{equation}%
We study the \ Ricci scalar depending on the scale factor from Eq. (\ref{evv}%
) and take the limit of Eq. (\ref{16r}) to show that the non-singular
curvature, the Ricci tensor, and the Kretschmann scalar when the universe
accelerates at $a(t)\rightarrow 0$ and at $a(t)\rightarrow \infty $. 
\begin{equation}
\lim_{a(t)\rightarrow 0}R(t)=\frac{24\kappa ^{2}}{l^{2}},
\end{equation}%
\begin{equation}
\lim_{a(t)\rightarrow 0}R_{\mu \nu }R^{\mu \nu }=\frac{144\kappa ^{4}}{l^{4}}%
,
\end{equation}%
\begin{equation}
\lim_{a(t)\rightarrow 0}R_{\mu \nu \alpha \beta }R^{\mu \nu \alpha \beta }=%
\frac{96\kappa ^{4}}{l^{4}},  \label{18}
\end{equation}%
\begin{equation}
\lim_{a(t)\rightarrow \infty }R(t)=\lim_{a(t)\rightarrow \infty }R_{\mu \nu
}R^{\mu \nu }=\lim_{a(t)\rightarrow \infty }R_{\mu \nu \alpha \beta }R^{\mu
\nu \alpha \beta }=0.
\end{equation}%
In the future, the acceleration of the universe will stop at an infinite time,
and then the spacetime will become flat, without any singularities. The critical
scale factor to show the boundary of the universe acceleration is obtained
using Eqs. (\ref{100}) and (\ref{evv}) as follows: $a(t)<a_{c}=\frac{1}{%
2^{7/12}}\frac{\sqrt{B_{0}}}{\sqrt[4]{\beta }}$. Hence, the acceleration of
the universe is zero at the critical value of the scale factor $a=a_{c}$,
and at the early time we show that the universe accelerates until the
critical scale factor $a=a_{c}$, which is suggested by the model describing inflation without singularities.

\section{Evolution of the Universe}

In this section, we study the dynamics of the universe by using Einstein's equations and the energy density given in Eq. (\ref{pp}). We are
interested in the early regime of the
universe without using the dustlike matter. First, with the help of the
second Friedmann equation, we find the evolution of the scale factor as
given by 
\begin{equation}
\left( \frac{\dot{a}}{a}\right) ^{2}=\frac{\kappa ^{2}\rho }{3}-\frac{%
\epsilon }{a^{2}},  \label{19}
\end{equation}%
where $\epsilon =0,+1$ and $-1$ depending on the geometry of the universe
(flat,closed and open respectively). Then, using the Eq. (\ref{Feqn}) the value of the critical scale
factor is obtained as $a_{c}=\frac{1}{2^{7/12}}\frac{\sqrt{B_{0}}}{\sqrt[4]{%
\beta }}$. Now, one shows that a cosmic time \cite{cosmotime} is calculated
by 
\begin{equation}
t=\frac{1}{3A}\,\left( {2}^{3/4}{C}^{3/4}{a}^{3}+3\,\ln \left( a\right)
\right) +t_{0},  \label{tt}
\end{equation}%
where $t_{0}$ is a constant of integration which gives only the shift in
time, $A=\frac{2\,{\kappa }^{2}}{l^{2}}$ and $C=\frac{\beta }{B_{0}^{2}}=%
\frac{1}{a_{c}^{4}}$. Note the assumptions that the universe is flat ($\epsilon =0$) and the integral constant is $t_{0}=0$. Furthermore, Eq. (\ref{tt}) can be written in the units of the critical scale factor as%
\begin{equation}
t=\frac{1}{3A}\,\left( {2}^{3/4}\frac{{a}^{3}}{a_{c}^{3}}+3\,\ln \left(
a\right) \right),   \label{tttt}
\end{equation}%
so one finds 
\begin{equation}
{a(t)=\mathrm{\exp }}\left[ -\frac{1}{3}\,\mathbf{LambertW}\left( \frac{{2}%
^{3/4}{\mathrm{e}^{3A\,t}}}{a_{c}^{3}}\right) +At\right] ,
\end{equation}%
and we consider $t=0$ to find the equation
for the radius of the universe 
\begin{equation}
\frac{1}{3A}\left( {2}^{3/4}\frac{{a}^{3}}{a_{c}^{3}}+3\,\ln \left( a\right)
\right) =0,
\end{equation}%
where the solution is found as 
\begin{equation}
a_{0}=a(t=0)={\mathrm{\exp }}\left[ -\frac{1}{3}\,\mathbf{LambertW}\left( 
\frac{\beta ^{3/4}{2}^{5/2}}{B_{0}^{3/2}}\right) \right] .  \label{phase}
\end{equation}%
The function of $a_{0}$ is a radius of the universe. Hence, the Eq.(\ref%
{phase}) represents the phase of the universe without any singularity $(t=0)$%
. The cosmic time calculated in Eq.(\ref{tttt}) has no singularity, and the
scale factor as a function of time has almost exponential behaviour as shown
in Fig.(3). It is noted that small cosmic time in the regime of the early
universe depends on the NMM fields and they play essential role for the
evolution of the universe in the early regime. This shows also that as $%
a_{0}\approx 0.77a_{c}<a_{c}$, the universe experiences accelerating
expansion, without need for dark energy models. The acceleration of the
universe begins at the initial radius of the universe $a_{0}$ and it stops
at the critical value $a_{c}$, where the acceleration of the universe is
zero $\ddot{a}=0$. After the acceleration stops, the universe decelerates
until the big crunch. 
\begin{figure}[tph]
\centering\includegraphics[width=8cm]{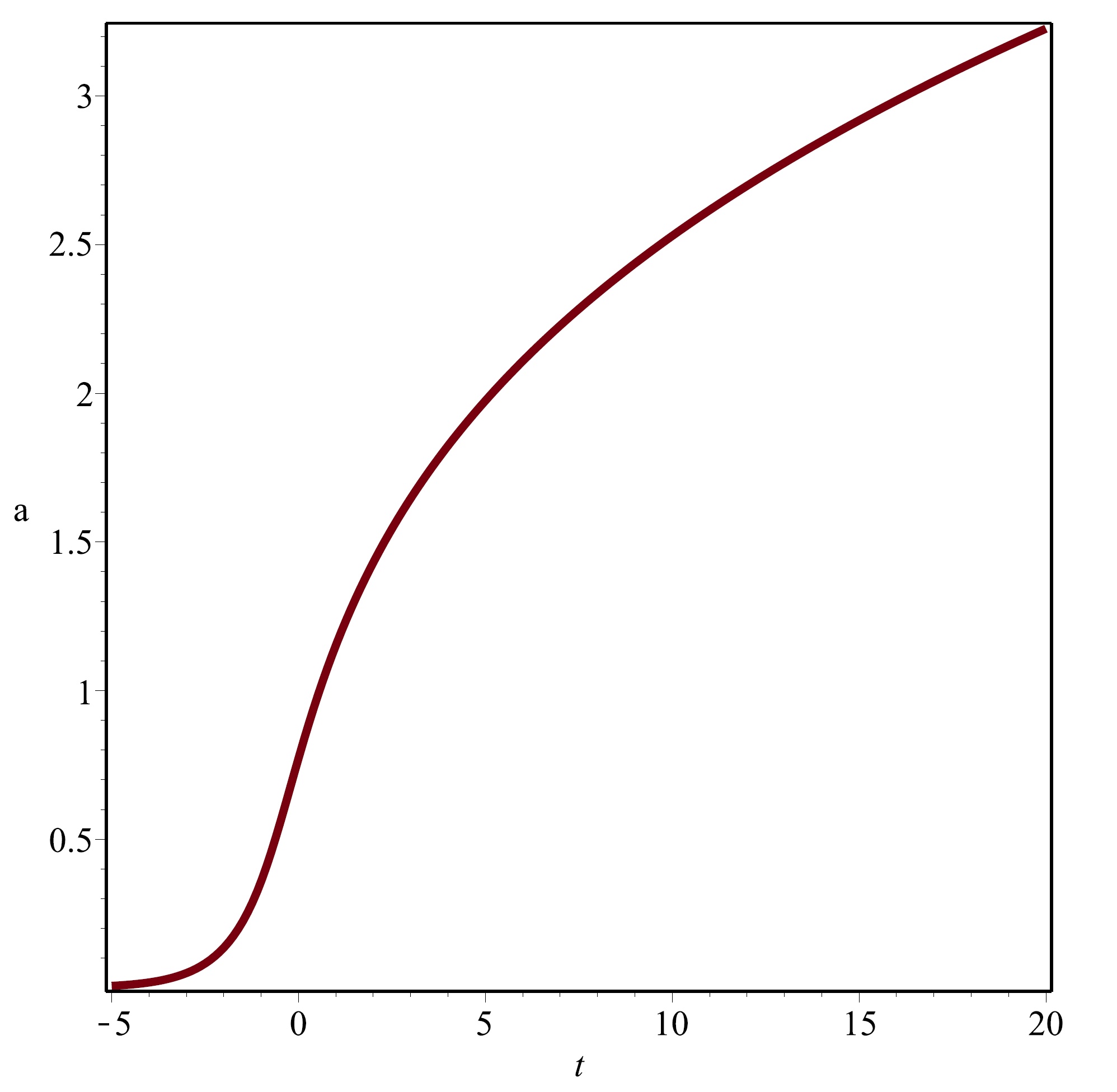}
\caption{plot of the scale factor $a$ versus the time $t$ (for $A=1$ and $C=1$)}
\end{figure}
Furthermore, for the positive time cases there is no singularities
for the spatial curvature $K=\frac{1}{a^{2}}$ when $t\rightarrow 0
$and $t\rightarrow \infty $, however when $t\rightarrow
-\infty $, clearly the scale of spatial curvature goes to infinity
leaving the closed universe, 
\begin{equation}
\lim_{t\rightarrow 0}K=\text{1.67},~~\lim_{t\rightarrow \infty }K=0,\text{ }%
\lim_{t\rightarrow -\infty }K=\infty .
\end{equation}

\subsection{A test of causality with Speed of the Sound}

A well-known way to test the causality of the universe in order to
survive, is by using the speed of the sound, which must be less than the local
light speed, $c_{s}\leq 1$ \cite{sound}$.$ The next requirement is \ based
on $c_{s}^{2}>0,$ positive value of the square sound speed. Those
requirements satisfy a classical stability. The square of the sound speed is
obtained \ from Eqs. (\ref{rho}) and (\ref{p}); 
\begin{equation}
c_{s}^{2}=\frac{dp}{d\rho }=\frac{dp/d\mathcal{F}}{d\rho /d\mathcal{F}}
\label{cs1}
\end{equation}%
\begin{equation}
=-\,\frac{\sqrt[4]{2}\left( 5\,\mathit{B}_{0}^{2}2^{3/4}\sqrt[4]{{\frac{%
\beta \,{a}^{4}}{\mathit{B}_{0}^{2}}}}+\sqrt{2}{a}^{4}\beta \right) {\frac{1%
}{\sqrt[4]{{\frac{\beta \,{a}^{4}}{\mathit{B}_{0}^{2}}}}}}}{3\mathit{B}%
_{0}^{2}\left( 1+{2}^{3/4}\left( {\frac{\beta \,{a}^{4}}{\mathit{B}_{0}^{2}}}%
\right) ^{3/4}\right) },
\end{equation}%
and the classical stability $(c_{s}^{2}>0)$ occurs at 
\begin{equation}
5\,\mathit{B}_{0}^{2}2^{3/4}\sqrt[4]{{\frac{\beta \,{a}^{4}}{\mathit{B}%
_{0}^{2}}}}+\sqrt{2}{a}^{4}\beta <0.  \label{cs2}
\end{equation}

The inequality of $c_{s}\leq 1$ is satisfied in Eq.(\ref{cs1}) for a positive valuse of $\beta $ ($\beta >0$). It should be noted that the magnetic field strengh $B_{0}$ can have any value. Then, to satisfy the classical
stability, in Eq. (\ref{cs2}) we find the limit of the scale factor as $%
a(t)>5^{1/3}2^{1/12}\frac{\sqrt{B_{0}}}{\sqrt[4]{\beta }}\simeq 1.81\frac{%
\sqrt{B_{0}}}{\sqrt[4]{\beta }}$. \ So the deceleration of the universe
occurs at this stage after reach the critical value of acceleration finished
at $a_{c}=\frac{1}{2^{7/12}}\frac{\sqrt{B_{0}}}{\sqrt[4]{\beta }}\simeq 0.67%
\frac{\sqrt{B_{0}}}{\sqrt[4]{\beta }}$. Therefore, a superluminal
fluctuation of the universe ( $c_{s}\geq 1$) does only occurs at the early
deceleration phase of the universe in the cosmological model with NMM
fields. Furthermore, this model has a classical instability at $a(t)<1.81\frac{\sqrt{B_{0}}}{\sqrt[4]{\beta }}$.
This instability can be explained by the inflation period and the short universe deceleration time, a result of the uncontrollable growth of the
energy density perturbation.

\section{Conclusion}

In this paper, we used the model of NMM fields with parameters $\beta $ and $l$ for the sources of the gravitational field. This model is not scale-invariant because of the free parameters $\beta $ and $l$, so the
energy-momentum tensor is not zero. We consider the universe to be magnetic and to accelerate with the help of NMM field sources. After the inflation period, it was shown that the universe is homogeneous and
isotropic. The acceleration of the universe is bounded at $a(t)<a_{c}(t)=%
\frac{1}{2^{7/12}}\frac{\sqrt{B_{0}}}{\sqrt[4]{\beta }}$.

We also showed that, at the time of the Big Bang, there was no singularity in the energy density, pressure, or curvature terms. After some time, the universe approaches flat spacetime. We checked causality and found that it satisfies the classical stability where the speed of sound should be
less than the local light speed. Hence, nonlinear sources, such as NMM fields
at the early regime of the universe, allow accelerated expansion with inflation and without dark energy. This model of NMM fields can also be used
to describe the evolution of the universe. We noted that at the
weak NMM field, there is no Maxwell's limit, so the inflation and the
acceleration of the universe can be analyzed by using different types of
fields. We manage to smooth the singularity of the magnetic universe by using only the NMM fields that were strong in the accelerated phase of the universe. In our model, in the early regime of the universe, NMM fields are very strong, making the effects of the usual electromagnetic fields negligible. We leave for a future publication the use of NMM fields with the usual Maxwell fields, and scalar fields, to investigate this problem more deeply. Another future project is to find the relationship between the different
types of NMM fields and the possible existence of wormholes in the universe
and their effect of Hawking radiation in relation to our previous work 
\cite{worm1,hr0,worm2,hr1}. 

\section{Acknowledgment}

The author would like to thank Prof. Dr. Mustafa Halilsoy for reading the manuscript and giving valuable suggestions. 
The author is grateful to the editor and anonymous referees for their valuable and constructive suggestions.


\begin{thebibliography}{99}
\bibitem{supernova1} S. Capoziello and V. Faraoni, Beyond Einstein Gravity:
A Survey of Gravitational Theories for Cosmology and Astrophysics (Springer
Science+Business Media\ B.V.,\ New York, 2011).

\bibitem{sing} B. Craps, arXiv:1001.4367 [hep-th].

\bibitem{muk} F.\ V. Mukhanov, et al. Phys.Rev.Lett.\textbf{\ 68}, 1969-1972
(1992).

\bibitem{cos1} E. N. Saridakis, M. Tsoukalas, arXiv:1601.06734 [gr-qc].

\bibitem{cos2} H. Sheikhahmadi, E.N. Saridakis, A. Aghamohammadi, K. Saaidi,
arXiv:1603.03883 [gr-qc].

\bibitem{born1} Garcia-Salcedo, Ricardo et al. Int.J.Mod.Phys. A\textbf{15},
4341-4354 (2000).

\bibitem{camara} C.S. Camara, et al. Phys.Rev. D \textbf{69}, 123504 (2004).

\bibitem{eliz} E. Elizalde,et al. Phys.Lett. B \textbf{574}, 1-7 \ (2003).

\bibitem{coupling} C. Quercellini , M. Bruni, A. Balbi, D. Pietrobon,
Phys.Rev. D \textbf{78}, 063527 (2008).

\bibitem{novello0} V. A. De Lorenci, R. Klippert, M. Novello, J. M. Salim,
Phys.Rev. D \textbf{65}, 063501 (2002).

\bibitem{novello1} M. Novello, et al. Phys.Rev. D \textbf{69}, 127301 (2004).

\bibitem{novello2} M. Novello, et al. Class.Quant.Grav. \textbf{24},
3021-3036 (2007).

\bibitem{vol1} D. N.Vollick, Phys.Rev. D \textbf{78, }063524 (2008).

\bibitem{ak3} O. Akarsu, T. Dereli, Int.J.Theor.Phys. \textbf{51}, 612-621
(2012).

\bibitem{beck} A.W. Beckwith, J.Phys.Conf.Ser. \textbf{626,} no.1, 012058 \
(2015).

\bibitem{kruglov1} S. I. Kruglov, Phys.Rev. D \textbf{92}, no.12, 123523
(2015).

\bibitem{kruglov2} S. I. Kruglov, Int. J. Mod. Phys. D \textbf{25}, no. 4,
1640002 (2016).

\bibitem{genci} G. Gecim, Y.Sucu, arXiv:1603.00352 [gr-qc].

\bibitem{nonm1} Y.Cai , E.N. Saridakis, M. R. Setare, J. Xia, Phys.Rept. 
\textbf{493}, 1-60 (2010).

\bibitem{nonm2} E.N. Saridakis, S. V. Sushkov,\ Phys.Rev. D \textbf{81},
083510 (2010).

\bibitem{horava} E.N. Saridakis, Eur.Phys.J. C \textbf{67}, 229-235 (2010).

\bibitem{ak1} O. Akarsu, T. Dereli, N. Oflaz, Class.Quant.Grav. \textbf{32},
no.21, 215009 (2015).

\bibitem{const} N. Breton, R. Lazkoz, A. Montiel, JCAP \textbf{1210}, 013
(2012).

\bibitem{novello3} M. Novello, S.E.P. Bergliaffa, \ Phys.Rept. \textbf{463},
127-213 (2008).

\bibitem{novello4} V.F. Antunes, M. Novello, Grav.Cosmol.\textbf{\ 22},
no.1, 1-9 (2016).

\bibitem{novello5} E. Bittencourt, U. Moschella, M. Novello, J.D. Toniato,
Phys.Rev. D \textbf{90}, no.12, 123540 (2014).

\bibitem{ak2} O. Akarsu, T. Dereli, N. Katirci, M. B. Sheftel, Gen.Rel.Grav. 
\textbf{47}, no.5, 61 (2015).

\bibitem{born} M. Born, L. Infeld, Proc.Roy.Soc.Lond. A \textbf{144},
425-451 (1934).

\bibitem{magneticU1} R. Durrer, A. Neronov, Astron Astrophys Rev \textbf{21}%
, 62, (2013).

\bibitem{w} K. E. Kunze, Plasma Phys.Control.Fusion \textbf{55}, 124026
(2013).

\bibitem{cosmotime} A. Balbi, EPJ Web Conf. \textbf{58}, 02004 (2013).

\bibitem{NNM} M. Halilsoy, A. Ovgun, S.Habib Mazharimousavi, Eur.Phys.J. C 
\textbf{74}, 2796 (2014).\ 

\bibitem{worm1} A. Ovgun, M. Halilsoy, Astrophys Space Sci (2016) 361:214.

\bibitem{hr0} I. Sakalli, A. Ovgun, Eur. Phys. J. Plus \textbf{131}: 184
(2016).

\bibitem{worm2} A. Ovgun, I. Sakalli, Theoretical and Mathematical Physics, 190(1): 120-129 (2017).

\bibitem{hr1} A. Ovgun, Int.J.Theor.Phys. \textbf{55}, 2919 (2016).

\bibitem{tolman} R. Tolman, P. Ehrenfest, Phys.Rev. \textbf{36}, no.12, 1791
(1930).

\bibitem{sound} R. Garcia-Salcedo, T. Gonzalez, I. Quiros, Phys.Rev. D 
\textbf{89}, no.8, 084047 (2014).
\end{thebibliography}
\end{document}